\documentclass[twocolumn,pra,showpacs,amsmath,amssymb,nofootinbib]{revtex4}
\usepackage{amsthm,url}

\begin{document}
\title{Reply To ``Comment on `Quantum Convolutional Error-Correcting Codes'
 ''}
\author{H. F. Chau}
\email{hfchau@hkusua.hku.hk}
\affiliation{Department of Physics, University of Hong Kong, Pokfulam Road,
 Hong Kong}
\affiliation{Center of Theoretical and Computational Physics, University of
 Hong Kong, Pokfulam Road, Hong Kong}
\date{\today}

\begin{abstract}
 In their comment, de Almedia and Palazzo \cite{comment} discovered an error in
 my earlier paper concerning the construction of quantum convolutional codes
 \cite{convolution}. This error can be repaired by modifying the method of code
 construction.
\end{abstract}

\pacs{03.67.Pp, 89.70.+c}
\maketitle
 de Almedia and Palazzo \cite{comment} found a counter-example showing the
 invalidity of Theorems~2 and~3 in my earlier paper in Ref.~\cite{convolution}.
 Their counter-example is correct; and the source of error lies with the proof
 of Lemma~2 in Ref.~\cite{convolution}. In fact, Lemma~2 is not correct and
 Theorem~3 should be modified as follows. (It is straight-forward to extend the
 modified theorem to cover the case of qudits.)

\begin*{\par\medskip\noindent {\bf Theorem~3.~}}
 Let $C_1$ be a classical (block or convolutional) code of rate $r_1$ and
 distance $d_1$ and let $C_2$ be the $[n_2 = d_2,1,d_2]$ majority vote
 classical code of rate $r_2$, namely, the one that maps $|t\rangle$ to
 $\bigotimes_{j=1}^{n_2} |t\rangle$. We construct a quantum code $C$ by first
 encoding a quantum state by $C_1$, then by applying a Hadamard transform to
 every resultant qubit, and finally by encoding each of the Hadamard
 transformed qubit by $C_2$. The rate and minimum distance of code $C$ equal
 $r_1 r_2$ and $\min (d_1^\perp,d_2)$ respectively, where $d_1^\perp$ is the
 minimum distance of the (classical) dual code of $C_1$.
\end*{}
\begin{proof}
 Clearly, the rate of code $C$ equals $r_1 r_2$. So, we only need to show that
 its minimum distance is $\min (d_1^\perp,d_2)$. Let us examine the classical
 code $C_1$ and the quantum code $C$ in the stabilizer formalism. We denote the
 operation of applying $\sigma_x$ ($\sigma_z$) to the $i$th qubit by $X_i$
 ($Z_i$). The encoded operation that flips the spin of the $i$th unencoded
 qubit for the classical code $C_1$ can be expressed in the form $X_1^{f_i(1)}
 \circ X_2^{f_i(2)} \circ \cdots \equiv \prod_{j\geq 1} X_j^{f_i(j)}$, where
 $f_i$ is a binary-valued function. The dual code of $C_1$ is a linear space
 spanned by vectors in the form $\prod_{j\geq 1} X_j^{g_s(j)}$, where $g_s$'s
 are some binary-valued functions. Since $C_2$ is the majority vote code, from
 our construction of $C$, the encoded operation that flips the spin (shifts the
 phase) of the $i$th unencoded qubit for the quantum code $C$ is given by
 $\prod_{j\geq 1} \prod_{k=1}^{n_2} Z_{n_2 (j-1) + k}^{f_i(j)}$ ($\prod_{j\geq
 1} \prod_{k=1}^{n_2} X_{n_2 (j-1) + k}^{f_i(j)}$). Furthermore, the stabilizer
 of $C$ equals the span of $\{ Z_{n_2 (m-1) + 1} \circ Z_{n_2 (m-1) + \ell},
 \prod_{j\geq 1} \prod_{k = 1}^{n_2} X_{n_2 (j-1) + k}^{g_s(j)} : \ell = 2,3,
 \ldots ,n_2 \mbox{~and~} m,s\geq 1 \}$. So just like CCS codes, the spin flip
 and phase shift errors in the quantum code $C$ can be corrected separately.
 After explicitly writing down the encoded operations and the generators of the
 stabilizer for the (degenerate) quantum code $C$, it is straight-forward to
 check that $C$ detects all spin errors happened to less than $d_2$ qubits.
 Moreover, all phase shift errors involving with less than $\min (d_1^\perp,
 d_2)$ qubits are in the stabilizer. Thus, the minimum distance of code $C$ is
 $\min (d_1^\perp ,d_2)$.
\end{proof}

\begin{acknowledgments}
 This work is supported by the RGC grant HKU~7010/04P of the HKSAR Government.
\end{acknowledgments}

\bibliography{qc34.1}
\end{document}